\documentstyle[aps,twocolumn]{REVTeX}

\input {epsf}

\newcommand{\beq}{\begin{equation}}
\newcommand{\eeq}{\end{equation}}
\newcommand{\beqa}{\begin{eqnarray}}
\newcommand{\eeqa}{\end{eqnarray}}

\def\opone{\leavevmode\hbox{\small1\kern-3.8pt\normalsize1}}

\begin{document}

\title{Experimental Test of Relativistic Quantum State Collapse with Moving
Reference Frames}
\author{H. Zbinden, J. Brendel, W. Tittel and N. Gisin \\
{\small {\em Group of Applied Physics, University of Geneva, 1211 Geneva 4,
Switzerland}}}
\date{\today}
\maketitle


\begin{abstract}
An experimental test of relativistic wave-packet collapse is
presented. The tested model assumes that the collapse takes place in the reference frame
determined by the massive measuring detectors. 
Entangled photons are measured at 10 km distance within a time interval of less than 
5 ps. The two apparatuses are in relative motion so that both detectors, each in
its own inertial reference frame, are first to perform the measurement. The data
always reproduces the quantum correlations and thus rule out a class of collapse models. 
The results also set a lower bound on the ''speed of
quantum information'' to $\frac{2}{3}10^{7}$ and $\frac{3}{2}10^4$ times the speed of light in the Geneva and the
background radiation reference frames, respectively.
The very difficult and deep question of where the collapse takes place - if it
takes place at all - is considered in a concrete experimental context. 
\end{abstract}


Entanglement, one of the most important features of quantum mechanics,
is at the core of the famous Einstein-Bohr philosophical debate 
\cite{eprBohr} and is the principal resource for Quantum Information Processing 
\cite{QIPIntro98}. The tension between
quantum mechanics and relativity has already received quite a
lot of attention in the context of local hidden variables and Bell inequality.
There the idea was to complete quantum mechanics with additional variables
that would reduce it to a classical theory. Here the intuition is that
this tension could be a guide for new physics, beyond quantum mechanics.

Despite the lack of a completely loophole free test of Bell inequality, the vast majority of
physicists is convinced that quantum mechanics correctly describes the atomic
world, including the correlation between distant systems, and we fully support
this conclusion. Yet, some physicists would like to treat the state
vector $\psi$ as describing an objective reality and not merely the physicists
information. We feel that such an approach is of interest, especially when it offers new experimental tests.
The difficulty of realistic interpretations comes from the "wave packet collapse" and there
seem to be only two alternatives. The first one assumes that the collapse
 is only a relative
phenomenon: the observer, the measuring apparatus and the quantum system under test all get
correlated in a way described by the Schr\"odinger equation such that all future
observations are consistent. In this description there is no real random choice, rather
all outcomes happen in different worlds that are in quantum superposition
(a so-called many-world or relative state interpretation \`a la Everett \cite{Everett}). The
second alternative assumes a real collapse with a real objective choice. It is
generally believed that there is no observable difference between these alternatives.
However, this conclusion depends on the exact form of the postulated collapse. For
example, the GRW \cite{GRW86} and the Primary State Diffusion \cite{Percival94}
models predict tiny differences, though these are not measurable with today's technology. 

In this letter we test the assumption of real collapse in two ways. First, we follow the idea
put forward by Suarez and Scarani that the collapse takes place 
in the reference frame determined by the measuring apparatus \cite{SuarezScarani97}. 
Next, we set experimental limits on the speed of the collapse, both in the 
reference frame defined by the massive environment of the experiment, that is in the
Geneva reference frame, and in the reference frame determined by the cosmic background
radiation.

Let us elaborate on the intuition of Suarez and Scarani. 
Each measuring apparatus defines a reference frame that we call measuring frame. 
In each measuring frame some  measurements (performed on distant systems) happen
before this one, while some happen later. The assumption is that the probabilities of outcomes
are determined by the local quantum state (as in standard quantum mechanics) and that
the local quantum state is collapsed by all the measurements that happened 
before in this measuring frame. If there is only one measuring frame, then this is
identical to quantum mechanics with the projection postulate, hence the predictions
are indistinguishable from those of standard quantum mechanics. In particular, they are
compatible with all previous experiments.
However, when applied to EPR-like situations
with two distant and moving measuring apparatuses, defining different reference frames\cite{spacelike}, 
special relativity implies that
the chronology of the measurements can differ from
one measuring frame to another. Let us discuss the case
of two entangled distant systems that for convenience we attribute to Alice and Bob. 
Assume first that in both measuring frames Alice's measurement takes place before Bob's.
In such a case Alice produces first an outcome with probabilities determined by her 
local state (obtained
by tracing over Bob's system). Next, Bob produces an outcome with probabilities also
determined by his local state, but this state takes into account Alice's result (i.e.
Bob's state is collapsed). This is quite common reasoning. Assume next
that Alice and Bob are in relative motion such that in each of the two measuring frames
the local measurement takes place before the distant one. In this case the probabilities 
of each result are determined by the local state without any collapse and models
\`a la Suarez-Scarani predict no correlations at all.
This is in strong
opposition to the quantum mechanical predictions according to which the correlation should
be observed independently of any time ordering.
{The case that both measurements take place after the distant one is discussed in 
\cite{suarezlong}}. 

Let us emphasize how natural our tests are in the context of assumed 
real collapses. Indeed, since the collapse is non local, 
if it is a real physical phenomenon it must happen in
some privileged frame. It is then natural to assume that the latter is either defined 
by the measuring apparatus, or by the massive environment, or by the background radiation 
field \cite{CBR}. The attractive aspect of these ideas is that they lead to difficult but
feasible experiments which could severely reduce the room for "real collapse models". 
They also set the question of what is a measurement in a concrete
context, since the measuring device determines the relevant measuring frame \cite{meas}.

The general idea of collapse models \`a la Suarez-Scarani is clear. However, in order
to design feasible tests, a more specific model has to be elaborated. Indeed, to test the
general idea an entire measurement apparatus would have to be put in relativistic motion.
Fortunately, in any collapse model there is an
assumed intrinsic irreversible choice after which the collapse has happened. 
It thus suffices to speed up the device where this happens. We call this
device the {\it choice device}.
In their original work, Suarez and Scarani \cite{SuarezScarani97} assumed that the
beam-splitters are the choice devices, inspired by Bohm's model.
The experiment described in this letter tests the more conventional assumption that collapses
are produced by all detectors and absorbing materials. The motivation for the latter
is that the relevant physics in detectors happens
in the first layers where the irreversible absorption takes place
in less than a picosecond. Note that negative results, i.e. particle not detected or
not absorbed, do also produce a collapse. In summary, the first detector or absorber encountered by 
any particle acts as the {\it choice device}. In a binary choice, as in our experiment, when
a particle encounters a second detector or absorber in the absolute future of the 
{\it choice device}, then the collapse already happened and 
there is no longer any alternative: the second device merely
reveals the choice made by the {\it choice device}.

In order to test the above model, we took
advantage of our long-distance Bell-experiment presented earlier in more detail 
\cite{Tittel98,Tittel99}, see Fig. 1. A source of energy-time entangled photons is 
located in a Swisscom terminal in the center of Geneva. The photons are sent through 
the optical fiber telecom network to two villages, Bellevue and Bernex separated by 
10.6 km. There, the photons are analyzed by two identically imbalanced fiber optic 
Michelson interferometers. The cases when the photons either both take the short arm 
of their interferometers, or both take the long arm, are indistinguishable, 
leading to interference according to Feynman's criterion. 
One interferometer is kept 
at a constant temperature, while the temperature of the other one is scanned, producing a phase 
variation. 
We can thus continuously measure the correlation as a function of the phase.

According to special relativity the time ordering of the measurements differ between
the two measuring frames only if their relative speed $v$ and their time difference 
$\delta t$ satisfy:
\begin{equation}
\delta t\leq \frac{Lv}{c^{2}}
\label{vmin}
\end{equation}
where $L=10.6$ km is the straight line distance between Alice and Bob. Consequently,
assuming a speed $v$ of 100 m/s and a safety margin of 2, a $\delta t\approx5$ ps 
timing accuracy is necessary. Note that achieving this automatically provides a bound
on the speed of the assumed collapse. Indeed, if both measurements take place before the
quantum information from the distant measurement result reaches them,
then no correlation would be observed. 

The technical challenges are thus, first to achieve a speed of 100 m/s, next to
adjust the fibers relative lengths below 5 ps (corresponding to 1 mm, each fiber 
being about 10 km long),
third to master the dispersion such that the pulse spreading remains below 5 ps.

To achieve an absorber speed of 105m/s we use a 20 cm diameter black-painted
aluminium disk of 1 cm thickness directly driven by a brushless
250W DC motor (Maxon EC) turning vertically at 10000 rpm.
During the absorption, the circular motion provides a good 
approximation to a linear one, defining thus the inertial reference frame.
It is oriented with a compass to make it run away 
or towards the other observer. 

We now describe the fiber lengths adjustment and dispersion management.
To adjust roughly the
distances from the source to the detectors we add 1.5 km of 
optical fiber on a spool on the link to Bellevue.
Furthermore, in order to equilibrate the chromatic dispersions, 
we add about 500 meters of dispersion shifted fiber on the link
to Bernex, this is necessary because the dispersion of this link
is higher (dispersion shifted fibers have relatively high negative
dispersion around 1300 nm). Finally, each fiber link measures in total about
10 km and has about 9 dB losses. Next, the fiber lengths are measured with a
home made OTDR (Optical Time Domain Reflectometer) with a precision of a few
cm. Short fiber pigtails are used for this adjustment. In a further step, we
use another home made low coherence interferometer \cite{PMDint} 
with 100 $\mu m$ resolution (using a LED with a 2 nm FWHM
interference filter). Fine tuning is realized by pulling on a 2 meter long
fiber on a rail with a micrometer-screw (optical fibers have 1\% elasticity).

The dispersion induced spreading of the wavepackets may easily be larger
than the achievable precision of the positioning. To limit the spreading we
take advantage of the 2-photon chromatic dispersion cancellation phenomena 
\cite{2photonCD,2photonCDbis}: if the central wavelength of the
downconverted photons is precisely at the (average) zero chromatic
dispersion wavelength of the two fibers, then, in the domain where chromatic
dispersion varies linearly, both photons undergo exactly the same delay. We
thus measured accurately and equilibrated the chromatic dispersion of the fiber links and
found the zero chromatic dispersion wavelength with a precision of $\pm 0.2$
nm. This uncertainty together with non-linear chromatic dispersion determines
the width of the 2-photon wave-packet.
Reducing the bandwidth of the downconverted photon with a 10 nm filter, we
estimate that the resulting spread is below 5 ps (for a detailed analysis, see
\cite{suarezlong}). 

Due to daily thermal expansion, the optical lengths between Geneva and each of the
villages changes by several mm over day. We observe that Bernex is drifting
further away during the daytime, since this link is more exposed to
temperature variations. The drift proves to be monotonic, in one direction
during the day and in the other one during the night. Accordingly, we aligned the
paths taking in account the daily drift such that the optical distances from
the source to the two interferometers will be perfectly equal some time
later. During these few hours we continuously record the 2-photon
interference fringes, scanning the phase of the Bellevue
interferometer. After an acquisition we confirm with a second measurement
that the path lengths really passed through the equilibrium point. In this way
we are sure that some fringes are collected when the fiber lengths difference
is smaller than 1 mm (corresponding to 5 ps). Many
interferograms were collected over various day and night periods and
measurement times. Fig. 2 displays typical data taken over 6 hours while the
optical link to Bernex lengthened by 2 mm with respect to the one to
Bellevue and the wheel was rotating, such that both measurements where ''before the
other'' over almost the entire scan. 
Inevitably, the curves show high statistical fluctuations due to
the low count rate. In spite of this, one can state that the visibility of
the two photon interferogram remains constant. Especially, a reduced
visibility over a scan span of 5 ps, as predicted by the model under test, should easily be
noticed. After substraction of the 237$\pm 5$ cts/100s accidental
coincidences (which are caused by the well understood phenomenon of dark counts and
which we measured independently), the fit of Fig 2 shows a constant fringe visibility of 83\%
\cite{bellviol}.

We also measured interferograms corresponding to the ''after-after''
configuration by inversing the rotation of the wheel,
again with no evidence for a breakdown of the correlations.
We like to mention that with the detectors as ''choice device''
a breakdown of the correlations would allow to exploit ''non-locality'' for
superluminal communication by slightly adapting the setup \cite{suarezlong}.

Our results do also demonstrate quantum
correlation measurements quasi-simultaneous in the natural Geneva reference frame,
setting a conservative, nevertheless impressive lower bound on the speed of
any hypothetical quantum influence, with $c$ the speed of light: 

\begin{equation}
\frac{10.6~km}{5~ps}\approx 2\cdot 10^{15}~\frac{m}{s}=\frac{2}{3}10^{7}~c,
\end{equation}

This speed remains superluminal in all reference
frames, in particular in that defined by the Cosmic Background Radiation (CBR).
Our earth moves at a mean speed of 371 km/s relative to this single-out frame.
Due to the earth rotations, both around its axis and around the sun,
the date and time when the data were taken and the orientation of the experiment
are relevant to establish the corresponding bound. A detailed analysis will be presented
elsewhere \cite{Scarani00}. Taking into account the Bellevue-Bernex orientation (almost exactly
north-south), the worst case correspond to a delay of 37 ns in the CBR frame and we
we get a conservative bound close to $10^3$c. 
We also performed this experiment without 
the rotating wheel and with two aligned detectors (APD 1 \& 2 on Fig.1). 
Note that for this bound, the 5 ps timing in the Geneva reference frame is unnecessary 
since anyway the timing in the background radiation frame is much less precise. Hence,
we could relax the constraint on the spectral width of the photons, accepting larger
dispersion and performed this experiment without 
the rotating wheel and with two aligned detectors (APD 1 \& 2 on Fig.1). 
This provides much higher counting rates, improves the signal to noise ratio and
sets a bound of $1.5\cdot10^4$c to the speed of quantum information
in the CBR frame \cite{Scarani00}.

Let us emphasize again that the above bound on the speed of ''quantum
information'' (quantum state preparation) is not in conflict with
relativity. What can be said is that Alice can predict with certainty the
quantum state of a photon 10 km away which was still in a completely mixed
state some ps before. Whether this implies the transmission of some kind of
information (or influence) is a matter of debate and models \cite{Eberhard89}. 
In this respect the recent progress on evaluating the cost of classical
communication for the simulation of quantum correlation is interesting \cite
{BCT99,Steiner99}: to simulate our experiment with classical communication
one would not only need superluminal communication, but the extreme
speed of 10 million time the speed of light.

We presented results of two experiments that go beyond the standard
tests of Bell inequality. Their objectives are to explore experimentally the
possible limits of quantum mechanics, to test the most
peculiar predictions of quantum physics and to open the road for further
experimental investigations of these no-longer purely philosophical
questions. Our
results fully support the quantum predictions, re-enforcing our confidence
in the possibility to base future understanding of our world and future
technology on quantum principles. 
They also contributes to the renewed
interest for experimental challenges to the interpretation of quantum
mechanics. ''Experimental
metaphysics'' questions \cite{expmetaph} like ''what about the concept of
states?'', ''the concept of causalities?'' will have to be (re)considered
taking into account the results presented in this letter. For example, our
results make it more difficult to view the ''projection postulate'' as a
compact description of a real physical phenomenon \cite{Pearle85,HPA89,GRW86}.
However this is
only a first example of this new class of tests and further experiments are needed before
general conclusions can be made. 

\small
This work would not have been possible without the financial support of the
"Fondation Odier de psycho-physique". It also profited from support by Swisscom and the Swiss
National Science Foundation. We would like to thank A. Suarez and V. Scarani for very
stimulating discussions and H. Inamori for preparing work during his stay in
our lab.






\section*{\protect\normalsize Figure captions}
\begin{enumerate}
\item  {\normalsize Schematic of the experiments that consist of
a photon pair source and two analyzers separated by 10.6 km, see \cite{Tittel99}. 
The absorbing surface A and the rotating wheel are at equal distances from the 
source. The detectors APD3 and APD4 are connected with longer fibers such that
each photon meets first the absorber, next the detector. 
In a second experiment the absorbers are replaced by two photon counters APD1 and 
APD2, again at exactly the same distance.
We obtain typically 
2 kcts/s single count rates and a mean value of about 3 coincidences per second (incl. 
2 accidentals), for details, see text and \cite{suarezlong}.}

\item  {\normalsize 2-photon interference fringes measured over 6 hours,
each data point corresponds to a time interval of 100s. The difference of
the optical path lengths is is varying from - 8 to + 1.3 ps. Negative values
mean that the detections occurs first in Bernex in the Geneva-Bernex
reference frame. In the moving Bellevue reference frame the detections
happen first in Bellevue over the entire scan range, as indicated on the
upper time scale. Despite this different time ordering no reduced visibility
is observed.}
\end{enumerate}


\begin{thebibliography}{99}
\bibitem{spacelike} 
For space-like separated measurements there are always reference frames such that
in the first frame one measurement happens first, while in the second frame it is the other
measurement which is first, see e.g. \cite{AharonovAlbert81}. 
But generally these frames are arbitrary, contrary to models \`a la
Suarez-Scarani where the frames are naturally determined by massive devices.
\bibitem{bellviol}
This visibility is large enough for a violation of Bell's inequality. 
However, a violation of Bell
inequality is not necessary in our context, since we do not test local 
models, but, on the opposite, non-local collapse models.
\end{thebibliography}
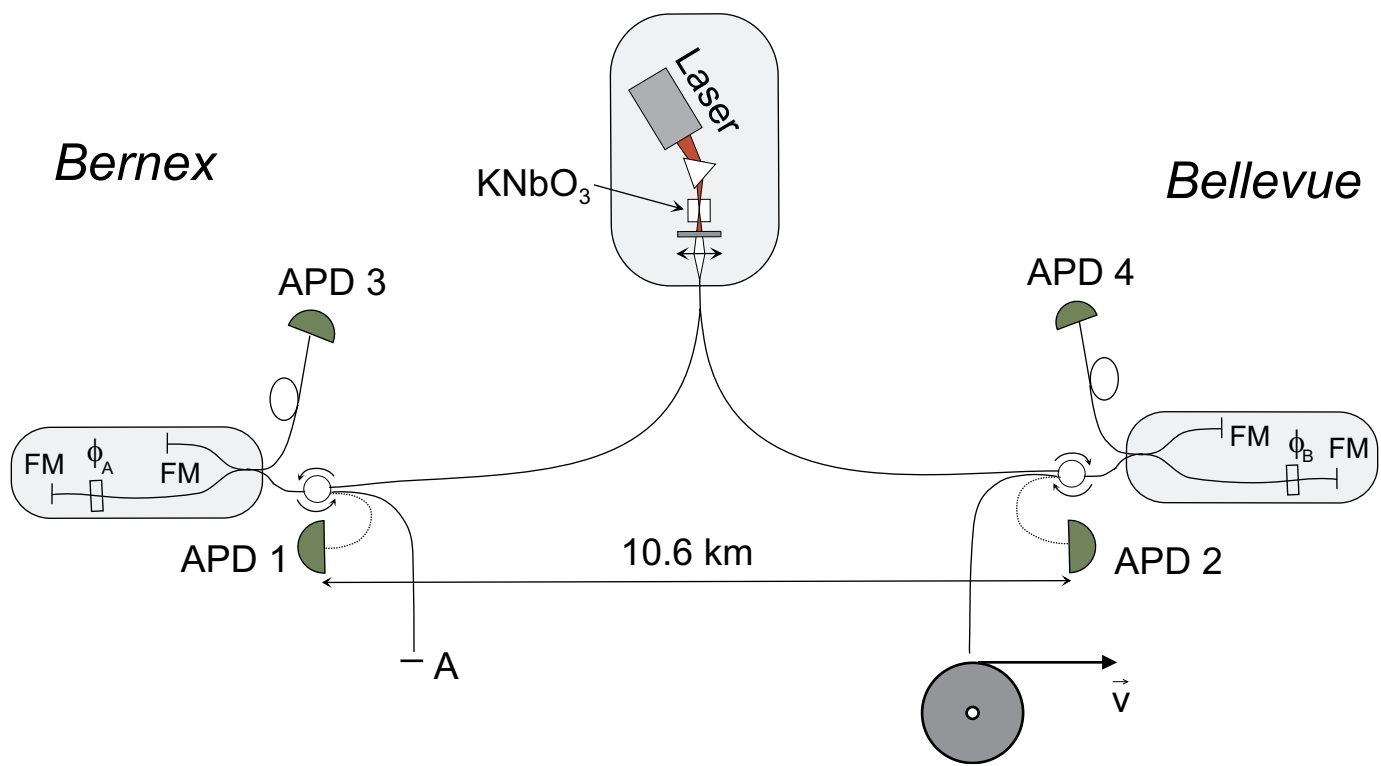
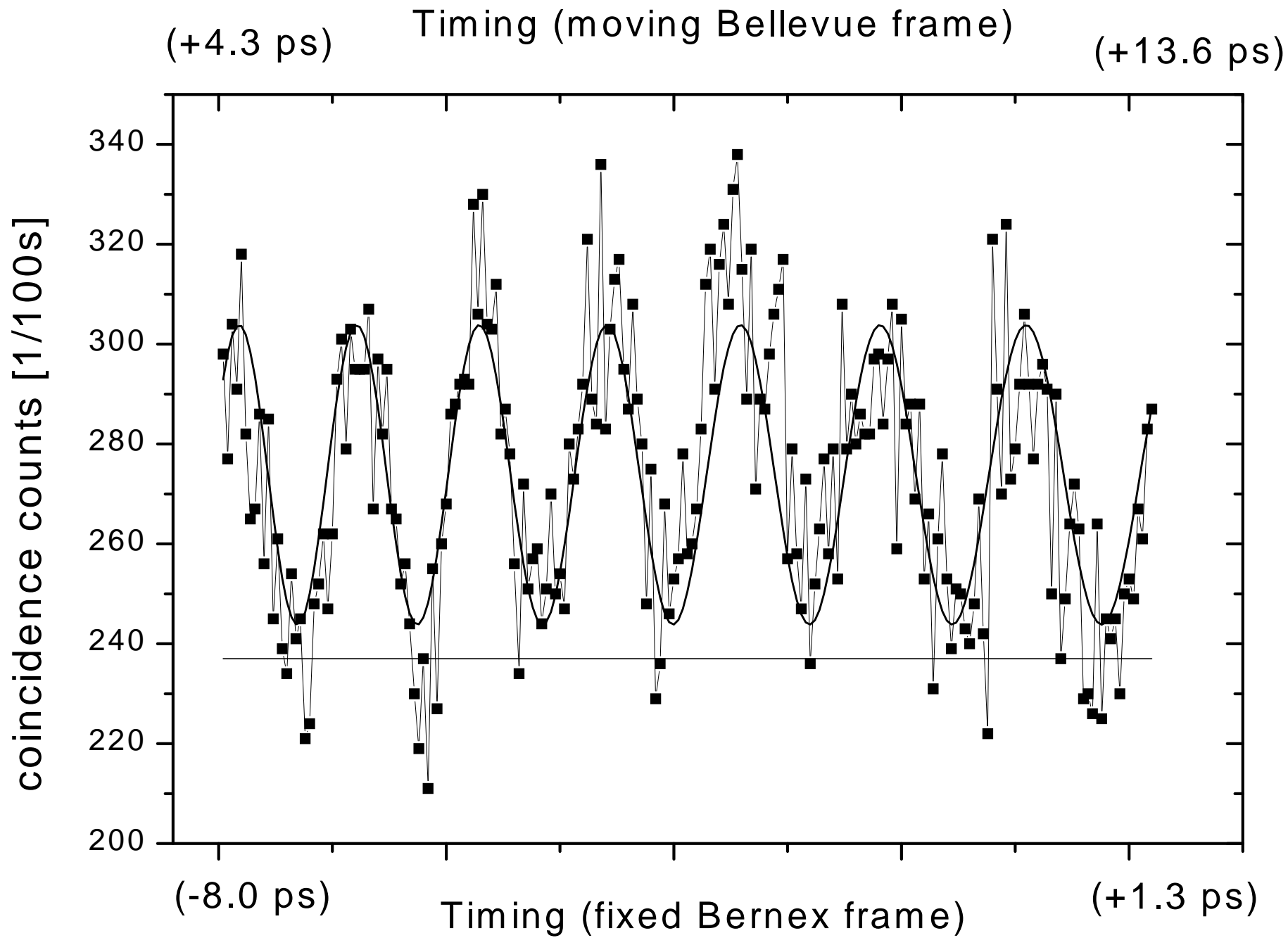
\end{document}